\documentstyle[aps,prl,epsfig]{revtex}

\begin{document}
\draft
\twocolumn[\hsize\textwidth\columnwidth\hsize\csname %
@twocolumnfalse\endcsname

\title{Inelastic tunneling through mesoscopic structures}
\author{Kristjan Haule$^a$ and Janez Bon\v ca$^{a,b}$}
\address{$^a$J. Stefan Institute, Ljubljana; $^b$FMF, University of 
Ljubljana, Ljubljana, Slovenia}
\date{\today}
\maketitle
\begin{abstract}\widetext
Our objective is to study resonant tunneling of an electron in the
presence of inelastic scattering by optical phonons.  Using a recently
developed technique, based on exact mapping of a many-body problem
onto a one-body problem, we compute transmission through a single site
at finite temperatures. We also compute current through a single site 
at finite temperatures and an arbitrary strength of the potential drop
over the tunneling region. Transmission vs. incident electron energy
at finite temperatures displays additional peaks due to phonon
absorption processes.  Current at a voltage bias smaller than the phonon
frequency is dominated by elastic processes. We apply the method to an
electron tunneling through the Aharonov-Bohm ring coupled to optical
phonons. Elastic part of electron-phonon scattering does not affect
the phase of the electron. Dephasing occurs only through inelastic
processes.
\end{abstract} 
] 

\narrowtext

\section{Introduction}
Advances in crystal-growth techniques and constantly shrinking
semiconductor devices have motivated researchers to study electron
tunneling in mesoscopic structures.  The ability to grow nearly
perfect microscopic devices has enabled experimental confirmation of
many theoretical predictions based on rather simple microscopic
models.  In particular, some basic ideas about electron-phonon
interaction motivated an explanation of  the seemingly unusual properties
of electrons tunneling in the presence of interactions with phonons
and other excitations \cite{tsui,kouw}, where inelastic processes
affect the peak-to-valley current ratio. This  is important in device
applications.  Another line of research in mesoscopic structures is
focused on Aharonov-Bohm oscillations in a mesoscopic
ring\cite{yacoby,buks,stern,yeyati,feng}. Special attention in this
field is devoted to a loss of coherence, or dephasing, under the
influence of inelastic scattering leading to suppression of $hc/e$
oscillations.

Most previous treatments of this problem use a Green's function
approach often based on Keldysh formalism
\cite{wingreen,glazman,gelfand,stoveng,jonson}.  Exact solutions were
obtained only in the wide-band limit and for certain special cases.
Calculating the current in the presence of inelastic scattering, some
authors work in the limit of large bias through the tunneling region.
This results in neglecting the backward current and the exclusion of
the filled final states \cite{wingreen}. Indepent-boson model has been
successfully used to directly solve the 1D Schr\" odinger equation for
arbitrary barrier structure \cite{cai}. Transfer-matrix approach has
proven very efficient to model realistic barrier structures and
compare calculations to experiment. Electron-phonon interaction is in this case
treated via the Fermi golden rule \cite{turley}.  Other attempts rely
on linear response theory using the Kubo formula
\cite{feng}.

This work is based on a recently developed method \cite{bonca} for
studying inelastic electron tunneling. The method provides exact
solutions for tunneling problems where a single electron tunnels in
the presence of phonon degrees of freedom that are limited to the
tunneling region. The main goals of this work are: a) to extend the
existing method to finite temperatures, b) to derive an approximate
formula for electric current in the presence of inelastic degrees of
freedom, and c) to apply developed formalism to the case of a single
site coupled to phonons as well as to  an Aharonov-Bohm ring. The method we
use allows numerically exact calculation of transmission at finite
temperatures.  We treat bands exactly in contrast to some previous
works \cite{wingreen,glazman}, where the wide-band limit was used. We
derive an approximate formula for the electric current that contains
an exact expression for the transmission matrix and provides results
for current at an arbitrary voltage drop through the tunneling region,
as precisely as the one-electron approximation allows. In deriving the
equation for current we treat the forward and backward current on an
equal footing. We also take into account exclusion from the filled
final states. However, we neglect effects of Coulomb interaction and
phonon mediated electron-electron interaction. Our method can be
applied to complicated tunneling structures containning
multiple-phonon degrees of freedom. In this work we present results for
two cases a) a single site coupled to optical phonons and b) to an
Aharonov-Bohm ring in a tight-binding approximation where each site is
coupled to Einstein phonons. We study the effect of inelastic
scattering on transmission and consequently the current through the
ring. We finally comment on the effect of inelastic scattering on
Aharonov-Bohm oscillations.

This work is organized as follows. In Sec.~II we introduce the
Hamiltonian for resonant tunneling in the presence of inelastic
degrees of freedom (optical polarons) that are limited to a small
region of space. We  describe the method and derive an
approximate equation for current.  In Sec.~III we present results for
transmission and current, at zero and finite temperature, through a
single site coupled to phonons. In Sec.~IV we give results for
transmission and current at zero temperature through the Aharonov-Bohm
ring. Concluding remarks and suggestions for future work are
presented in Sec.~V.

\section{Method}

The Hamiltonian we use can be written as a sum of the electron part
$H_{el}$, the phonon part $H_{ph}$ and finally the electron-phonon
interaction $H_{el-ph}$
\begin{eqnarray}
H &=& H_{el} + H_{ph} + H_{el-ph},\nonumber \\
H_{el} &=& \sum_j \epsilon _j c^\dagger _j c_j  -
\sum_{j,k} t_{j,k} ( c^\dagger _j c_k + h.c.),\nonumber\\
H_{ph} &=& \omega \sum_m   a^\dagger _m a_m ,\nonumber\\
H_{el-ph} &=& -  \sum_{j} \lambda_j c^\dagger _j c_j  
(a^\dagger _j + a_j).
\label{ham}
\end{eqnarray}
The potential, $\epsilon _j$ on site $j$, can describe a tunnel
barrier or a  voltage bias. Since we treat left and right lead exactly,
we choose a constant potential $\epsilon_{L(R)}$ within the left
(right) lead. The voltage drop is thus limited to the tunneling
region.  The hopping amplitude $t_{j,k}$ is set to $t$ within the
leads, $t_0$ is the hopping amplitude from the lead to the tunneling
region, and $\lambda_j$ is the (diagonal) coupling of an electron
on site $j$ to the phonon mode on the same site. Electron-phonon coupling is
limited to the tunneling region. We consider dispersionless Einstein
phonons with frequency $\omega$.

The problem we are facing is to solve a scattering problem of a single
electron in the presence of inelastic degrees of freedom. Since a
detailed explanation of the method was given in previous work
\cite{bonca}, we will present only a short overview that is necessary
for the reader to understand our generalization to finite temperature
and computation of the current.

\begin{figure}[tb]
\begin{center}
\epsfig{file=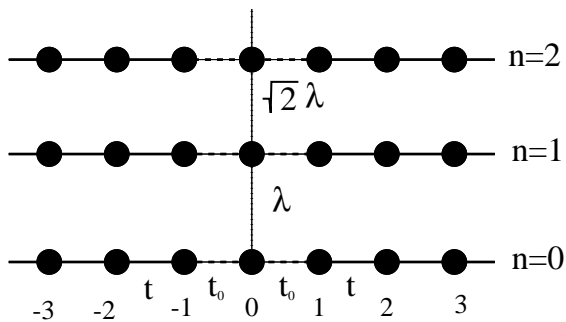,height=75mm,angle=-90}
\end{center}
\caption{ Each dot represents a basis state wavefunction
$\psi_{(j,n)}$ in the many-body Hilbert space.  The lowest row of dots
are the sites $j=-3,\dots,3$ with diagonal energies $\epsilon_j$.  The
rows above represent the same sites with $n=1,2$ phonon quanta on the
site $j=0$. Their diagonal energies are $\epsilon_j + \omega$ and
$\epsilon_j + 2 \omega$, respectively.  The bonds represent non-zero
off-diagonal matrix elements in the Hamiltonian.  The horizontal bonds
are the hopping amplitudes $t_{j,k}$.  The vertical bonds represent
the electron-phonon interaction.  The dots can also be interpreted as
Wannier orbitals in an equivalent 1-body tight-binding model.}
\label{tight}
\end{figure}
Consider for simplicity the case where the tunneling region consists of a
single site ($j=0$) with a single phonon mode that couples to the
electron density on site 0 (see the lowest row in Fig.~(\ref{tight})). 
Hopping matrix elements are $t_{k,l}=t$ for
the lead nearest neighbors $k,l\not=0$, and $t_{k,l}=t_0$ between site 0
and sites $\pm1$. The wavefunction can be written as $\psi_{(j,n)}$,
where the site index $j$ represents the position of the electron and $n$
represents the number of phonons on the site $j=0$.  The Schr\" odinger
equation can be written for this simple problem in the compact form, 
\begin{eqnarray}
E\psi_{(j,n)} &=& \epsilon_j \psi_{(j,n)} - \sum_{\langle
j,k\rangle}t_{jk} \psi_{(k,n)} + \omega n \psi_{(j,n)}\nonumber\\\
&-& \lambda
\delta_{j,0} \left(\sqrt{n+1} \psi_{(j,n+1)} + \sqrt{n}
\psi_{(j,n-1)}\right).  \label{schro} 
\end{eqnarray} 
It is already apparent from Eq.~(\ref{schro}) that a one-dimensional
many-body problem, consisting of an electron and different numbers of
phonon quanta, can be visualized as an effective two-dimensional
1-body problem with $n$ as the second dimension. For a better
perception we present Eq.~(\ref{schro}) graphically in
Fig.~(\ref{tight}) as a tight binding model in two dimensions. For
illustrative purposes we have restricted the variational space to a
maximum $N_{ph}=2$ phonon quanta.  At zero temperature, an electron
incident from the left is an incoming plane wave on the lower left
lead (there are no excited phonons on site $0$).  It has an amplitude
to exit on any of the six leads, corresponding to elastic and
inelastic backscattering and transmission. At finite temperatures a
finite number of phonon quanta $n$ are excited on site $j=0$ before
scattering with the probability
$P(n)=(1-e^{-\beta\omega})e^{-n\beta\omega}$. Thus an electron can
enter on any of the horizontal leads corresponding to different
$n$. We can solve the set of equations in Eq.~(\ref{schro}) taking
into account the boundary condition  specifying that an electron
can enter only through one lead at a time. Solutions within the leads
for an electron approaching from the left are:
\begin{eqnarray}
&&\ \psi_{(j<0,n)} = A^{(n)} \exp{(i k_L^{(n)} j)} + B^{(n)} 
\exp{(-i k_L^{(n)} j)}
\nonumber\\
&&\ \psi_{(j>0,m)} = C^{(m)} \exp{(i k_R^{(m)} j)} 
\nonumber\\
&&\ \psi_{(j<0,m \ne n)} = B^{(m)} \exp{(-i k_L^{(m)} j)},
\label{psi}
\end{eqnarray}
where $n(m)$ represents the number of excited phonon quanta before (after)
scattering, $A^{(n)}$ is the amplitude of the incident wave, and 
$B^{(m)}$ and $C^{(m)}$ represent reflection and transmission
amplitudes. Wavevectors are defined by the conservation of total
energy,
\begin{equation}
\epsilon_L+n\omega-2t\cos(k^{(n)}_L)=
\epsilon_{L(R)}+m\omega-2t\cos(k^{(m)}_{L(R)}). 
\label{energy}
\end{equation}
Using the pruning technique \cite{bonca} we can remove all the leads
that contain outgoing waves from the calculation. We are thus left
with a system of linear equations that connect wavefunctions on the
central site $\psi_{(0,n)}$ with the lead that carries the incoming
electron. This problem can be solved easily by recursion for
essentially any number of phonon quanta. The transmission matrix
$T_{L\to R}^{n,m}(\epsilon,\epsilon^\prime)$ is defined as
\cite{bonca}
\begin{equation}
T_{L\to R}^{(n,m)}(\epsilon,\epsilon^\prime) = 
\left|{C^{(m)} \over A^{(n)}} \right|^2
{\sin{k_R^{(m)}} \over \sin{k_L^{(n)}}},
\label{transm}
\end{equation}
where $n(m)$ represent incoming (outgoing) channels (by a channel we
denote the lead with a specified number of phonons ($n$)), and
$\epsilon$ ($\epsilon^\prime$) are incoming (outgoing) electron
energies counted from the middle of the band {\it i.e.}
$\epsilon=-2t\cos(k_L^{(n)})$.  We define the total and elastic
transmission as a sum over all incoming channels $n$, weighted by the
probability $P(n)$ and a sum over all outgoing channels. Since $P(n)$
depends on the temperature of the tunneling region, we must assume,
that the temperature of the tunneling region is well defined. This is
achieved by coupling it to the external heath-bath.
\begin{eqnarray}
T_{tot}(\epsilon)&=&\sum_{n,m}P(n)
T_{L\to R}^{(n,m)}(\epsilon,\epsilon^\prime), \label{total}\\
T_{elast}(\epsilon)&=&\sum_{n}P(n)
T_{L\to R}^{(n,n)}(\epsilon,\epsilon^\prime).
\label{elastic}
\end{eqnarray}
Using the presented technique we can compute transmission at zero and
finite temperatures with any desired accuracy. However, 
transmission is not in principle a directly measurable quantity. One
has to compute electric current in order to provide a measurable
quantity. The total transmission as defined in Eq.~(\ref{total}),
enters the equation for current only in the case of a high-voltage
bias where neither backward current nor the exclusion of filled states
in the right lead are taken into account \cite{wingreen}.
 
In deriving the equation for the current we start by observing that
inelastic processes can be viewed as multichannel electron
tunneling. The current from the left lead, due to an electron entering
the tunneling region through a channel $n$ and exiting through $m$,
is given by the integral over the incoming momenta $\hbar dk_L^{(n)}$
times the velocity of the incoming electron $v_L=(1/\hbar)
d\epsilon/dk_L^{(n)}$, times the transmission probability $T_{L\to
R}^{(n,m)}(\epsilon,\epsilon^\prime)$, times the appropriate
combination of Fermi functions from the left and the right lead. The
total current from the left is expressed as a sum over all channels
weighted with the appropriate Boltzmann factor.  The net current is
given finally by the difference between the right- and the left- flowing
currents

\begin{eqnarray}
J&=&{e\over \pi\hbar} \int_{-2t}^{2t} d\epsilon \sum_{n,m}P(n)
T^{(n,m)}_{L\to R}(\epsilon,\epsilon^\prime)f_L(\epsilon)
\left( 1-f_R(\epsilon^\prime)\right) \nonumber\\
&-&{e\over \pi\hbar}\int_{-2t}^{2t} d\epsilon^\prime \sum_{m,n}P(m)
T^{(m,n)}_{R\to L}(\epsilon^\prime,\epsilon)f_R(\epsilon^\prime)
\left( 1-f_L(\epsilon)\right),
\label{curr1}
\end{eqnarray}
where $T^{(m,n)}_{R\to L}$ is the transmission matrix for an electron
coming from the right.  Electron energies are constrained by the
energy conservation law, $\epsilon+\epsilon_L+n\omega=
\epsilon^\prime+\epsilon_R+m\omega$. Fermi functions $f_{L(R)}$
describe the left (right) lead. We can write equation for
current~(\ref{curr1}) in a more compact form, by taking  advantage
of the time-reversal symmetry under which $T^{(m,n)}_{R\to
L}(\epsilon^\prime,\epsilon)\to T^{(n,m)}_{L\to
R}(\epsilon,\epsilon^\prime)$ and the fact that the transmission
matrix is nonzero when both electron energies $\epsilon$ and
$\epsilon^\prime$  are within the band,
\begin{eqnarray}
J_{tot}&=&{e\over \pi\hbar} \int_{-2t}^{2t} d\epsilon 
\sum_{n,m} T^{(n,m)}_{L\to R}(\epsilon,\epsilon^\prime){\large [}
P(n) f_L(\epsilon)
\left( 1-f_R(\epsilon^\prime)\right) \nonumber\\
&-&P(m)f_R(\epsilon^\prime)
\left( 1-f_L(\epsilon)\right) {\large ]}.
\label{curr2}
\end{eqnarray}
We can also compute the elastic contribution to  the total current in 
Eq.~(\ref{curr2}) by imposing the constraint of elastic
tunneling  $n=m$
\begin{eqnarray}
J_{elast}&=&{e\over \pi\hbar} \int_{-2t}^{2t} d\epsilon 
\sum_{n} T^{(n,n)}_{L\to R}(\epsilon,\epsilon+\Delta\mu) P(n) \nonumber \\
& &\left [ f_L(\epsilon) - f_R(\epsilon + \Delta\mu)\right ],
\label{curre}
\end{eqnarray}
where $\Delta\mu = \epsilon_L-\epsilon_R$.  The total energy of the
electron $E=\epsilon+\epsilon_L=\epsilon^\prime+\epsilon_R$ is in this
case conserved.  In the limit of zero temperature Eq.~(\ref{curre})
gives the correct result for elastic tunneling (see {\it
e.g.}\cite{wingreen}).

Our derivation of equation~(\ref{curr2}) is based on the one-electron
approximation which leads to neglecting many-body effects as are a)
the Coulomb repulsion, b) phonon mediated electron-electron
interaction. Closely connected to the latter is our assumption, that
phonon distribution $P(n)$ is independent of electron current through
the tunneling region. Validity of the equation~(\ref{curr2}) is
therefore limited to cases, when the current is small and the
tunneling region is strongly coupled to the external heat bath. This
can be achieved even at large voltage bias when coupling to the leads
is weak, {\it i.e.} $t_0\ll t$.

\section{Results for a single site}

In our calculation we have set the hopping within the lead to $t=1$.
The maximum number of allowed phonon quanta $N_{ph}$ was selected such
that results fully converged for all chosen parameters of the system
at different temperatures.

\subsection{Transmission}
\begin{figure}[tb]
\begin{center}
\epsfig{file=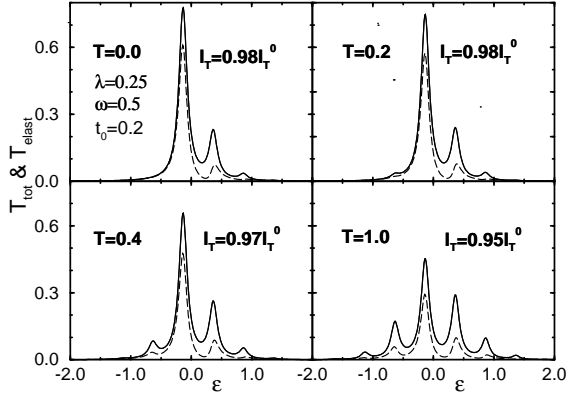,height=75mm,angle=-90}
\end{center}
\caption{ Transmission probability as a function of the incident
electron energy calculated for different temperatures $T$.  The heavy
line is the total transmission $T_{tot}$, and the dashed (lower) line
the elastic part $T_{elast}$. The parameters of the Hamiltonian are:
$\lambda=0.25$, $\epsilon_l=\epsilon_r=\epsilon_0=0.0$ (no voltage
drop across the dot), $t_0=0.2$, $\omega=0.5$ and $N_{ph}=17$.
Also presented are  sum-rules where $I_T^0=2\pi t (t_0/t)^2$.}
\label{trans}
\end{figure}

In Fig.~(\ref{trans}) we show the total and elastic transmission
(Eqs.~(\ref{total},\ref{elastic})) as a function of the incoming
electron energy $\epsilon$ at different temperatures. At low
temperatures we see a central peak positioned at $\epsilon \sim
\epsilon_0 -\lambda^2/\omega$, a one phonon side peak at $\epsilon
\sim \epsilon_0 + \omega -\lambda^2/\omega$, and a small two phonon
peak at higher $\epsilon$. Positions of the peaks approximately
correspond to polaron energies, given by $\epsilon_{pol}(n) =
\epsilon_0-\lambda^2/\omega + n \omega$; where, $n$ represents the
$n-$th excited state of a polaron. By polaron we symbolize a state of
an electron, coupled to phonon degrees of freedom located on the same
site in the limit when $t_0\to 0$. It is important to emphasize that
all peaks have an elastic and an inelastic contribution (see also
\cite{wingreen,bonca}). Elastic contributions correspond to tunneling
through the ground or the  excited state of a polaron without emitting (or
absorbing) a phonon. Inelastic parts correspond to tunneling through a
given state with simultaneous phonon emission or absorption. At zero
temperature only phonon emission processes are allowed since before
tunneling the phonon state contains no phonons.

At finite temperature other processes can take place. As an electron
enters the tunneling region there may be one or more phonons excited
in the system. The side peak at $\epsilon \sim \epsilon_0 -\omega
-\lambda^2/\omega$ is due to a process when an electron enters the
tunneling region and absorbs a phonon. Such a process has been observed 
by Cai {\it et al} \cite{cai}.
As the temperature increases
more inelastic channels open for electron tunneling, giving rise to an
increased strength of side peaks.  Interestingly enough, the sum-rule,
valid at zero temperature and within the wide-band approximation
\cite{wingreen,glazman} $I_T^0=\int d\epsilon T_{tot}(\epsilon)=2\pi
t(t_0/t)^2$, remains valid in our numerically exact approach at small
temperatures, and it changes at most 5\% at large temperatures (see
values in the inset).  We would like to stress, that sum-rules $I_T^0$
(and $I_J^0$ see next subsection), defined in the work of Wingreen
\cite{wingreen}, do not represent sum-rules in a strict sense.  They
instead represent identities that are derived on the basis of two main
approximations: dispersionless bands and a large bias limit used when
calculating the current. Even though our calculations are not limited
by these approximations, we nevertheless chose to compare integrals of
transmission with $I_T^0$ and in the next chapter integrals of current
with $I_J^0$.  As the temperature rises, peaks due to multi-phonon
processes increase in strength. It almost seems as though finite
temperature increases the effective coupling strength $\lambda$
\cite{jonson}. However, an increased coupling strength would also
shift the peaks. Peaks in Fig.~(\ref{trans}) do not shift with
increasing temperature. The increased strength of multi-phonon
side-peaks is a consequence of increased number of excited phonons in
the tunneling region at higher temperatures.
  
\subsection{Current through a single site}

\begin{figure}[tb]
\begin{center}
\epsfig{file=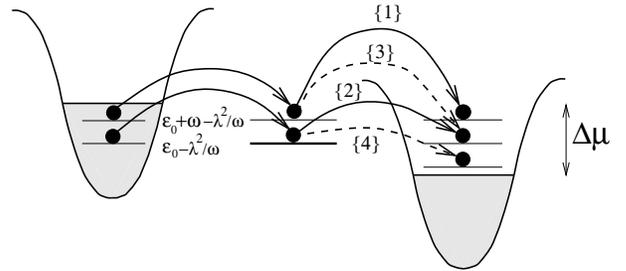,height=80mm,angle=-90}
\end{center}
\caption{ A schematic representation of the positioning of the bands.
Some elastic and inelastic processes that contribute to the total
current are shown. Shaded areas represent filled one-electron states.
In elastic processes (\{1\} and \{2\}) the total electron energy is
conserved $\epsilon+\epsilon_L = \epsilon^\prime+\epsilon_R$, while in
inelastic ones (\{3\} and \{4\}) it is not. For clarity, only zero- 
and one-phonon states on the central site are included.}
\label{band}
\end{figure}

We continue with a discussion of current. The relative positioning of
bands and the choice of chemical potentials in the calculation is
presented schematically in Fig.~(\ref{band}). We have positioned the
chemical potentials in the left and right lead in the middle of each
band.  This situation corresponds to both leads being made from the
same metal.  The value of the chemical potential within each lead is
constant and corresponds to asymptotic values in the leads.
Bands are shifted symmetrically due to an applied bias $\Delta\mu$
{\it i.e.} $\epsilon_L=\Delta\mu/2$ and $\epsilon_R=-\Delta\mu/2$,
thus $\epsilon_L-\epsilon_R=\Delta\mu=e V$; where, $V$ represents the
potential drop across the tunneling region.

\begin{figure}[tb]
\begin{center}
\epsfig{file=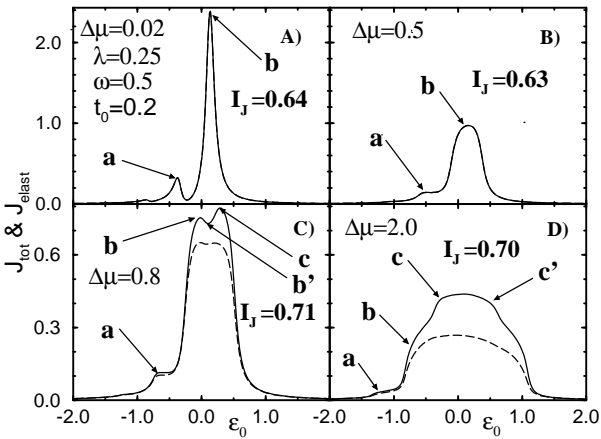,height=80mm,angle=-90}
\end{center}
\caption{ The total current $J_{tot}$ (solid line) and elastic current
$J_{elast}$ (dashed line) vs. $\epsilon_0$ at $T=0$ for four different
choices of $\Delta\mu$.  All curves are in units of $I_J^0$ (see
definition below).  Note also that the ordinates in figures A) and B) are
different than in C) and D).  Current flows at different $\epsilon_0$
through different channels.  Opening or closing of different channels
(processes) as $\epsilon_0$ increases is reflected in peaks, shoulders 
or dips in the curves.  Those anomalies (labeled by small letters)
at different $\Delta\mu$ correspond to the following processes (see
Fig.~(\ref{band})) 
A) a:\{1\}, b:\{2\}, B) a:\{1\}, b:\{2\}, C)
a:\{1\}, b:\{1,2,3\}, b$^\prime$:\{2\}, c:\{2,4\},
D) a:\{1\}, b:\{1,2,3\}, c:\{1,2,3,4\}
c$^\prime$:\{2,4\}.}
\label{curr}
\end{figure}

In Fig.~(\ref{curr}) we present the total and elastic current vs.
$\epsilon_0$ (the on-site energy of the site $j=0$) for four different
values of $\Delta\mu$. At a small $\Delta\mu=0.02\ll \omega$ there is
no inelastic contribution to the current (solid and dashed curves
overlap). The main reason for this interesting effect is that the
chemical potential difference across the tunneling region is smaller
than the minimum energy change $\omega$ necessary for the inelastic
tunneling. The necessary condition for an electron, with an incoming
energy $\epsilon$, to contribute to the current at $T=0$, besides
having a finite tunneling rate, can be expressed with two
inequalities: $\epsilon+\epsilon_L<\mu_L$, and
$\epsilon^\prime+\epsilon_R+m \omega>\mu_R$.  At small $\Delta\mu$
this condition leads to well defined peaks in the
$J_{tot,elast}(\epsilon_0)$ curves since only electrons with a total
energy in a narrow interval between the left and right chemical
potential can give rise to current. Therefore only processes labeled
by \{1\} and \{2\} in Fig.~(\ref{band}) contribute to formation of the
side (a) and the central peak (b) respectively.  We should stress that
phonons nevertheless still play an important role even at small voltage
biases.  The side-peak (a) at approximately
$\epsilon_0=\lambda^2/\omega-\omega$ represents an elastic process where
an electron elastically tunnels through the first excited state of a
polaron whose energy is $\epsilon_{pol}(1) = -\lambda^2/\omega +
\omega$, while the main peak (b) corresponds to elastic tunneling
through the polaron ground state $\epsilon_{pol}(0)$.

As $\Delta\mu$ increases to $\Delta\mu=\omega=0.5$ the side peak
develops into a shoulder and moves towards lower
$\epsilon_0\sim\omega^2/\lambda - \omega - \Delta\mu/2$ while the
central peak broadens.  This is a consequence of separating the left
and right chemical potentials and thus imposing less restrictive
conditions on the tunneling electron energies. In particular, 
the broadening of the
central peak  is caused by simultaneous tunneling through
processes labeled  \{1\} and \{2\} in Fig.~(\ref{band}); also note the
figure caption.  Inelastic contribution is still small.

The main features appearing at larger bias, {\it i.e} $\Delta\mu =
0.8$, are a) the emergence of the inelastic current (the solid and
dashed lines do not overlap) and b) development of a new structure in
the central peak region. The appearance of the inelastic current is
caused by the opening of the inelastic channels at
$\Delta\mu>\omega$. The new structure in the central peak region is
caused by opening and closing of elastic or inelastic tunneling
channels as $\epsilon_0$ increases [see caption to
Fig.~(\ref{curr})]. At even larger bias $\Delta\mu=4\,\omega=2.0$ the
current $J_{tot}(\epsilon_0)$ broadens even though the effects of
different channel contributions are still visible. The inelastic
current increases relatively to the elastic current.

We have also  computed integrals $I_J=\int d\epsilon_0
J(\epsilon_0)$ that according to Wingreen {\it et al.} \cite{wingreen}
should equal $I_J^0=(e/\pi\hbar)2\pi t(t_0/t)^2\Delta\mu$. Our
findings are that even though the current does scale with
$\Delta\mu$, nevertheless, our integrals deviate from sum-rules that
exist in the wide band and large bias limit. The main reason is that
the current flows through only selected channels allowed by the
difference in the chemical potentials and not through all the channels
as is the case in the limit of large bias and wide-bands.

\begin{figure}[tb]
\begin{center}
\epsfig{file=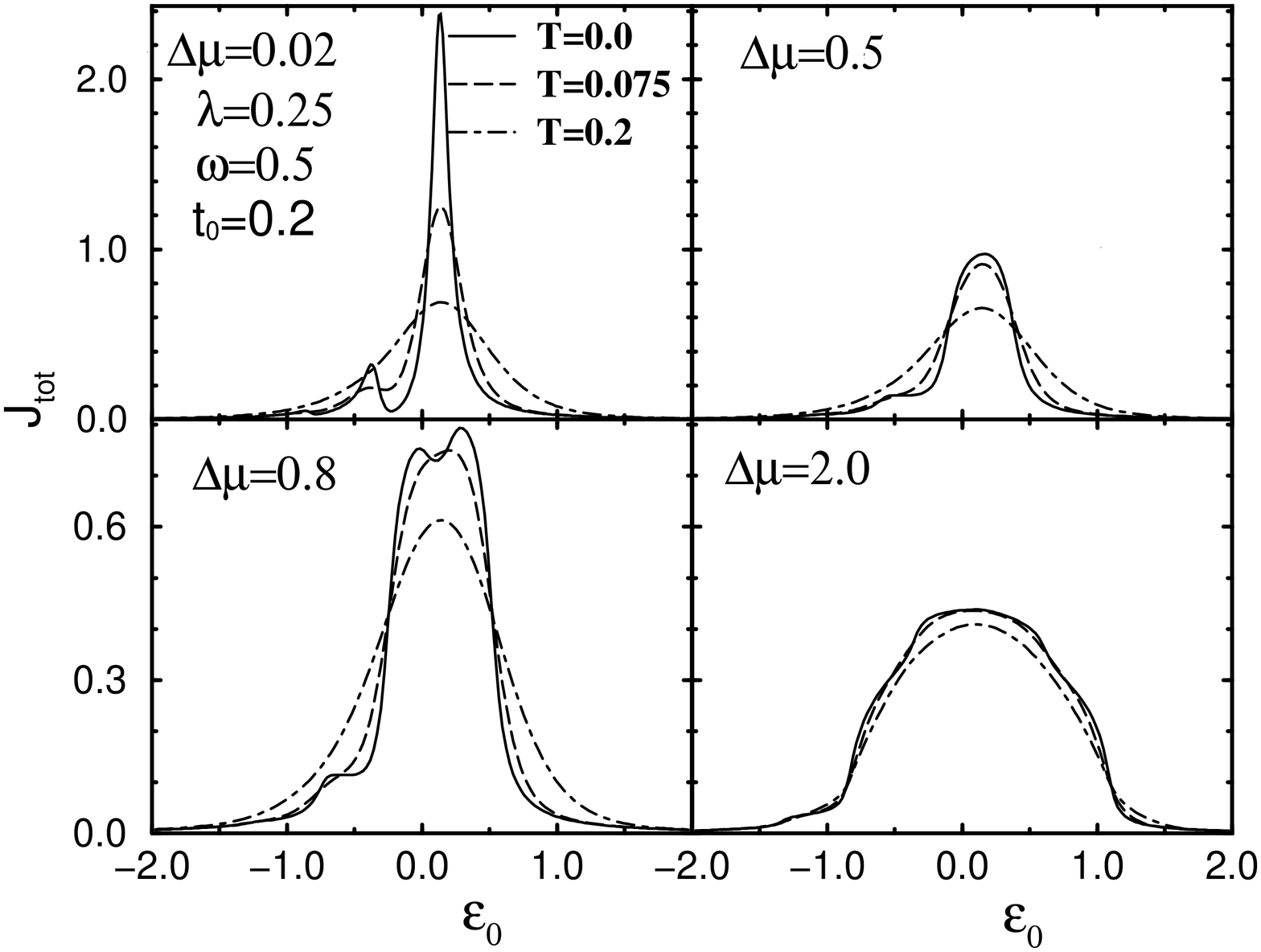,height=80mm,angle=-90}
\end{center}
\caption{ The total current $J_{tot}$ vs. $\epsilon_0$ at
$T=0,0.075$, and $0.2$ for four different choices of
$\Delta\mu$. All curves are in units of $I_J^0$. 
All other parameters are the same as in
Fig.~(\ref{curr}).}
\label{currt}
\end{figure}

The effect of finite temperatures on the total current is presented in
Fig.~(\ref{currt}).  In contrast to the case of the total transmission
we were unable to obtain additional peaks corresponding to phonon
absorption processes. With increasing temperature the
smearing by the Fermi functions overcomes development of those peaks.
The effects of raising the temperature are more intense at smaller
biases where peaks are narrower. Phonon side-peaks disappear around
$T\sim \omega/2$.

\section{Tunneling through Aharonov-Bohm ring}

Next, we consider tunneling and consequently the current through an
Aharonov-Bohm ring. The  purpose of this section is to investigate
the effect of dephasing by phonon modes on an electron as it tunnels
through a ring. In detail, we will focus on the effect of dephasing on
the transmission and current through the ring. 

\begin{figure}[tb]
\begin{center}
\epsfig{file=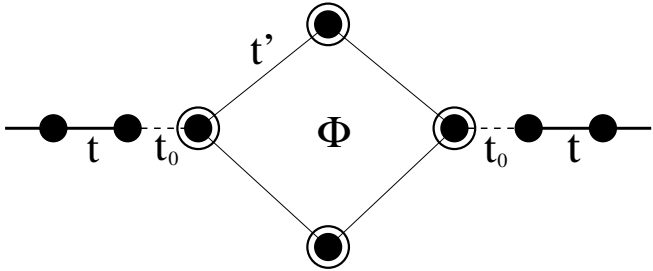,height=87mm,angle=-90}
\end{center}
\caption{Schematic representation of the Hamiltonian describing the
Aharonov-Bohm ring. Dots surrounded by circles represents sites
coupled to Einstein phonons. For simplicity we have chosen on-site
energies on the ring to be constant $\epsilon_j = \epsilon_0$. 
Magnetic flux $\Phi$ penetrates the center of the ring.}
\label{ring}
\end{figure}
The model is schematically presented in Fig.~(\ref{ring}). The ring
consists of four sites, connected by hopping matrix element
$t^\prime$, that are coupled to two connecting leads with the hopping
matrix element $t_0$.  Each site of the ring is coupled to an Einstein
phonon mode with frequency $\omega$. There are four different phonon
modes (one on each site) with identical frequency $\omega$. A
magnetic flux is penetrating the circle. This is reflected in an
additional phase $\pm\phi$ that the electron gains each time it hops
from one site on the circle to another. The electron part of the
Hamiltonian (\ref{ham}), describing hopping within the ring, has to be
modified in order to encompass the effect of a vector potential
\begin{equation}
H_{el}=\epsilon_0\sum_j c_j^\dagger c_j - t^\prime \sum_j\left (
e^{i\phi}c_j^\dagger c_{j+1} + h.c.\right ),
\label{hphi}
\end{equation}
where the sum runs over the sites of the ring. All on-site
energies within the ring were, for simplicity, set to $\epsilon_0$.  We
solve this problem using the  method  described above (see also
\cite{bonca}), and generalized to many phonon degrees of freedom. By
increasing the maximum allowed number of phonons $N_{ph}$ on each site
this problem can be solved with any desired accuracy. There are of
course computer limitations. In practice, for moderate electron-phonon
coupling and low temperatures, $N_{ph}=3$ is enough to obtain results
with at least 1\%  accuracy.  The number of many-body states
increases as $N_{st}=N (N_{ph}+1)^N=1024$, which corresponds to the
number of channels $N_{ch}=2N_{st}$.  $N$ is the number of sites in the
ring. In a scattering problem where the energy of the incoming
electron is known in advance, we have to solve a large $(N_{st}\times
N_{st})$ sparse system of complex linear equations for each
$\epsilon$. Due to computer limitations, we restricted
calculations to zero temperature.
\begin{figure}[tb]
\begin{center}
\epsfig{file=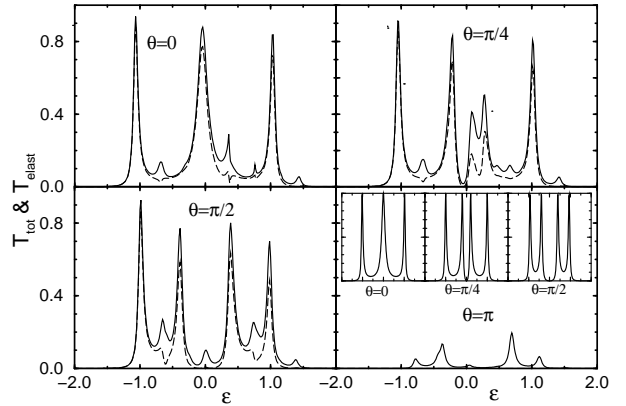,height=80mm,angle=-90}
\end{center}
\caption{ Transmission probability as a function of the incident
electron energy $\epsilon$ at $T=0$ calculated at different $\theta$.
The heavy line is the total transmission $T_{tot}$, and the dashed
(lower) line the elastic part $T_{elast}$. The parameters of the
Hamiltonian are: $\lambda=0.2$,
$\epsilon_l=\epsilon_r=\epsilon_0=0.0$, $t_0=0.3$, $\omega_0=0.4$
$t^\prime = 0.5$ and $N_{ph}=3$. In the inset we show 
(insets are in the same units as the rest of the figures in 
Fig.~(\ref{transa})) transmission
$T_{elast}$ for the noninteracting case, {\it i.e.} $\lambda=0$.}
\label{transa}
\end{figure}

We have computed the total and elastic transmission vs. $\epsilon$
through an Aharonov-Bohm ring at different values of total phase,
$\theta = \phi N$. In Fig.~(\ref{transa}) we show results for the
total and elastic transmission through a ring. Positions of the main
peaks are approximately located close to energies corresponding to
solutions of a tight binding problem on a $N=4$ site ring with a phase
$\phi$, {\it i.e.}  $E(k) = -2t^\prime \cos (k + \phi)$ for $k=2\pi
i/N$. One and two phonon side peaks (shoulders) are also visible at
approximately $\epsilon_{n,k}= E(k)+n\omega$.  The height and the width
of the main peaks change substantially when  the position
of the main peak coincides with the position of the phonon-side peak
(see $\theta=\pi/4$).

The most interesting result is found at $\theta=\pi$ which corresponds
to a phase, where in the case of noninteracting tunneling due to
negative interference, transmission is exactly zero for any
$\epsilon$. At finite electron phonon coupling we obtain, for
$\theta=\pi$, a finite total transmission. However, the elastic part
of transmission remains exactly (down to numerical accuracy)
zero. This  result is surprising considering that elastic tunneling
processes are possible when an electron first emits and then reabsorbs
a phonon. A phase shift that the electron undergoes while emitting a
phonon is thus exactly canceled by the phase shift after reabsorbing a
phonon. Tunneling at $\theta=\pi$ is therefore exclusively due to
inelastic processes. This finding is in accordance with the  well
accepted fact that impurity (elastic) scattering can not destroy
Aharonov-Bohm oscillations \cite{wash}, since it does not cause phase
decoherence.
\begin{figure}[tb]
\begin{center}
\epsfig{file=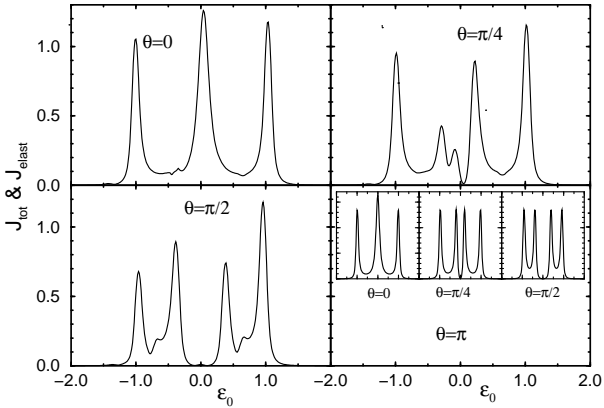,height=80mm,angle=-90}
\end{center}
\caption{ The total current $J_{tot}$ (solid line) and elastic current
$J_{elast}$ (dashed line) calculated at $\Delta\mu=0.1$
vs. $\epsilon_0$ at $T=0$ for four different choices of phase
$\theta$. All curves are in units of $I_J^0$. Note that the solid and
the dashed lines overlap indicating $J_{tot}=J_{elast}$. As in the
previous section we set $\epsilon_L=\Delta\mu/2$ and
$\epsilon_R=-\Delta\mu/2$.  The rest of the parameters remain the same as
in Fig.~(\ref{transa}).  In the inset we show for comparison the
elastic current $J_{elast}$ for the noninteracting case. Insets are
in the same units as the rest of the Fig.~(\ref{curra}).}
\label{curra}
\end{figure}

Furthermore, we have calculated current, Eqs.~(\ref{curr2})
and~(\ref{curre}), vs. $\epsilon_0$ through the Aharonov-Bohm ring at
small bias, $\Delta\mu=0.1$, zero temperature and different
$\theta$. These results are presented in Fig.~(\ref{curra}). As we
have already shown in the previous section, only elastic current can
flow at small bias. Nevertheless, effects of electron-phonon
interaction are clearly visible. There is no net current at
$\theta=\pi$ since only the elastic part of transmission can
contribute to the elastic current.  We predict that at small bias
$\Delta\mu\ll \omega$ Aharonov-Bohm oscillations through the
mesoscopic Aharonov-Bohm ring should not diminish significantly due
to electron-phonon coupling, where coupling is limited to optical
phonons.

\begin{figure}[tb]
\begin{center}
\epsfig{file=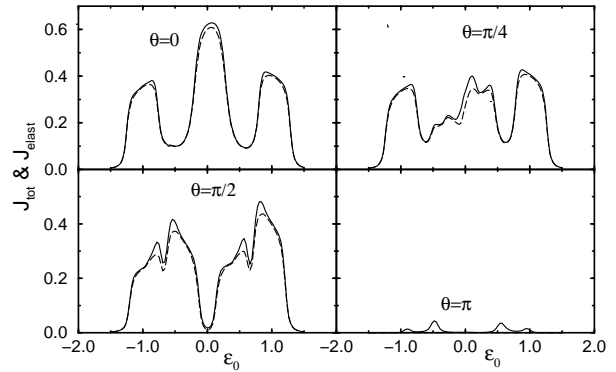,height=80mm,angle=-90}
\end{center}
\caption{ The total current $J_{tot}$ (solid line) and elastic current
$J_{elast}$ (dashed line) calculated at $\Delta\mu=0.5$
vs. $\epsilon_0$ at $T=0$ for four different choices of phase
$\theta$. The rest is the same as in Fig.~(\ref{curra}).}
\label{curra1}
\end{figure}

Last, we present results for current through the Aharonov-Bohm ring at
large bias, $\Delta\mu=0.5>\omega$. Results are presented in
Fig.~(\ref{curra1}). Two most important effects of larger bias are
broadening of the peaks and the appearance of the inelastic current
($J_{tot}-J_{elast}$). As a cosnequence of dephasing by inelastic
scattering we  observe finite inelastic current at $\theta=\pi$. 

\section{Conclusions}

In summary, we have extended a numerically exact method \cite{bonca},
for inelastic tunneling, to calculate transmission through a single
site coupled to phonons at finite temperatures. We have further
proposed an approximate formula for current based on exact results for
the transmission matrix, taking into account filled Fermi seas left
and right from the tunneling region and left and right flowing
currents. Presented formalism gives correct result in the limit of
elastic current. Our approach is entirely based on a one-electron
approximation. By investigating tunneling through an Aharonov-Bohm
ring coupled to optic phonons, we have obtained numerically exact
results for transmission at zero temperature. We have also computed
the current through the ring.

 We highlight some of the important findings of this work:

{Transmission through a single site vs. incident electron energy at
finite temperatures displays additional peaks due to phonon absorption
processes. The sum rule $I_T$, valid in the wide band limit and zero
temperatures, remains approximately obeyed in our exact approach even
at finite temperatures. Deviations are within 5\% even at temperatures
$T>\omega$.}
  
{Current through a single site mimics transmission curve at
small bias $\Delta\mu\ll \omega$ and $T=0$. At  voltage bias 
$\Delta\mu<\omega$, and zero temperature only current due to elastic
processes is possible, since inelastic processes are excluded due to
filled final states. With increasing temperature features due to
inelastic processes disappear around $T\sim\omega/2$.  Phonon
absorption peaks do not appear in the current at finite temperatures
since they are smeared by Fermi functions.}

{Transmission through the Aharonov-Bohm ring consists only of
inelastic transmission at $\theta=\pi$. The lack of elastic
transmission is considered as evidence that elastic processes, even
though they are a part of inelastic electron-phonon scattering, do not
change the phase of the tunneling electron.  We have thus presented
numerical proof that elastic scattering by phonons does not affect
Aharononv-Bohm oscillations. Dephasing occurs only through inelastic
processes. This is  the reason why interference
effects, measured at small bias $\Delta\mu<\omega$ across the
tunneling region, display strong Aharonov-Bohm oscillations by
changing the flux through the ring. Calculation of current at larger bias 
$\Delta\mu>\omega$ shows that only inelastic current flows through the
Aharonov Bohm ring in the case when $\theta=\pi$. The voltage
drop over the region has to be larger than the minimal energy difference
necessary for inelastic scattering {\it i.e.} $\Delta \mu > \omega$.}
    
  It would be interesting to study the tunneling through the
Aharonov-Bohm ring at finite temperatures. Such computation would
demand larger Hilbert space and more computational time. Our method
also allows the implementation of the nonlinear electron-phonon
interaction and linear and nonlinear phonon-phonon interactions. It
would be fascinating to investigate, how phonon-phonon interactions,
leading to internal phonon dynamics, affect dephasing of the tunneling
electrons through an Aharonov-Bohm ring.

\section{Acknowledgements}

 One of the authors (J.B.) gratefully acknowledges support by the SLO
USA grant and the hospitality of Los Alamos National Laboratory. We
thank Anton Ram\v sak, Peter Prelov\v sek and Stuart Trugman for
stimulating discussions and Mike Stout for careful reading of the
manuscript.



\end{document}